\documentclass[11pt]{article}
\usepackage{graphicx}
\usepackage{epsf}  

\newcommand{\psfile}[3][]{ 
  \begin{center}
    \setlength{\epsfxsize}{#3\linewidth}\leavevmode
    \def\noOpt{}\def\testit{#1}\ifx\testit\noOpt%
      \epsfbox{#2}%
    \else%
      \epsfbox[#1]{#2}%
    \fi
  \end{center} 
}

\newcommand{\psfiletwoBB}[5]{ 
  \begin{minipage}{\linewidth}
    \parbox[b]{.49\linewidth}{%
      \begin{center}
        \setlength{\epsfxsize}{#5\linewidth}\leavevmode\epsfbox[#1]{#2}
      \end{center}
    }
    \hfill
    \parbox[b]{.49\linewidth}{%
      \begin{center}
        \setlength{\epsfxsize}{#5\linewidth}\leavevmode\epsfbox[#3]{#4}
      \end{center}
    }
  \end{minipage}
}

\newcommand{\BABARPubYear}    {03}

\newcommand{\BABARConfNumber} {009}
\newcommand{\SLACPubNumber} {9692}

\input pubboard/babarsym
\newcommand{\calB}{\mbox{${\cal B}$}}

\def\dbline{\noalign{\vskip 0.15truecm\hrule}\noalign{\vskip 2pt}\noalign{\hrule\vskip 0.15truecm}}
\def\sgline{\noalign{\vskip 0.20truecm\hrule\vskip 0.20truecm}}

\newcommand{\etapKst}{\mbox{$B\ra\etapr K^*$}}
\newcommand{\etappi}{\mbox{$B\ra\etapr\pi$}}
\newcommand{\etap}{\mbox{$B\ra\eta\pi$}}

\newcommand{\etaK}{\mbox{$B\ra\eta K$}}

\newcommand{\etaKp}{\mbox{$B^+\ra\eta K^+$}}
\newcommand{\etaPip}{\mbox{$B^+\ra\eta \pi^+$}}
\newcommand{\etaKst}{\mbox{$B\ra\eta K^*$}}
\newcommand{\etapK}{\ensuremath{B\ra\eta^\prime K}}

\newcommand{\etaKz}{\ensuremath{\Bz\ra\eta K^0}}

\newcommand{\fetaggk}{\ensuremath{\eta_{\gamma\gamma} K^{+}}}
\newcommand{\fetaggkz}{\ensuremath{\eta_{\gamma\gamma} K^{0}}}
\newcommand{\fetaggpi}{\ensuremath{\eta_{\gamma\gamma} \pi^{+}}}

\newcommand{\DE}{\ensuremath{\Delta E}}
\newcommand{\xf}{\mbox{${\cal F}$}}

\newcommand{\costhr}{\ensuremath{\cos\theta_{\rm T}}}

\newcommand{\kzs}{\mbox{$K^0_S$}}

\newcommand{\UfourS}{\mbox{$\Upsilon(4S)$}}

\newcommand{\pvec}{{\bf p}}
\newcommand{\half}{\mbox{${1\over2}$}}

\def\babar{{\em B}{\footnotesize\em A}{\em B}{\footnotesize\em AR}}

\newcommand{\acp}{\ensuremath{\calA_{ch}}}

\newcommand{\fetapp}{\ensuremath{\eta\pi^+}}
\newcommand{\etapp}{\ensuremath{\Bp\ra\fetapp}}
\newcommand{\Betapp}{\ensuremath{\calB(\etapp)}}

\newcommand{\fetapip}{\ensuremath{\eta \pi^+}}
\newcommand{\etapip}{\ensuremath{\Bp\ra\fetapip}}
\newcommand{\Betapip}{\ensuremath{\calB(\etapip)}}
\newcommand{\retapip}{\ensuremath{4.2^{+1.0}_{-0.9}\pm 0.3}}
\newcommand{\Retapip}{\ensuremath{(\retapip)\times 10^{-6}}}
\newcommand{\Aetapip}{\ensuremath{-0.51^{+0.20}_{-0.18}\pm 0.01}}
\newcommand{\fetakp}{\ensuremath{\eta K^+}}
\newcommand{\etakp}{\ensuremath{\Bp\ra\fetakp}}
\newcommand{\Betakp}{\ensuremath{\calB(\etakp)}}
\newcommand{\retakp}{\ensuremath{2.8^{+0.8}_{-0.7}\pm 0.2}}
\newcommand{\Retakp}{\ensuremath{(\retakp)\times 10^{-6}}}

\newcommand{\Aetakp}{\ensuremath{-0.32^{+0.22}_{-0.18}\pm 0.01}}
\newcommand{\fetakz}{\ensuremath{\eta\Kz}}
\newcommand{\etakz}{\ensuremath{\Bz\ra\fetakz}}
\newcommand{\Betakz}{\ensuremath{\calB(\etakz)}}
\newcommand{\retakz}{\ensuremath{2.6^{+0.9}_{-0.8}\pm 0.2}}
\newcommand{\Retakz}{\ensuremath{(\retakz)\times 10^{-6}}}
\newcommand{\uetakz}{\ensuremath{4.6}}
\newcommand{\Uetakz}{\ensuremath{\uetakz\times 10^{-6}}}

\newcommand{\Dcontrol}{\ensuremath{B^- \ra \pim \Dz,\ \Dz \ra K^{-} \pip \piz}}

\newcommand{\fetathrpipi}{\ensuremath{\eta_{3\pi} \pi^+}}
\newcommand{\fetathrpik}{\ensuremath{\eta_{3\pi} K^+}}
\newcommand{\fetathrpikz}{\ensuremath{\eta_{3\pi} K^{0}}}

\newcommand{\etahp}{\ensuremath{\Bp\ra\eta h^+}}

\newcommand{\etagg}{\ensuremath{\eta_{\gaga}}}
\newcommand{\etathrpi}{\ensuremath{\eta_{3\pi}}}
\newcommand{\etatogg}{\ensuremath{\eta\ra\gaga}}
\newcommand{\etatothrpi}{\ensuremath{\eta\ra\pi^+\pi^-\pi^0}}

\newcommand{\etal}{{\em et al.}}

\long\def\inst#1{\par\nobreak\kern 4pt\nobreak
    {\it #1}\par\vskip 10pt plus 3pt minus 3pt}

\begin{document}
{\pagestyle{empty}

\begin{flushright}
\babar-CONF-\BABARPubYear/\BABARConfNumber \\
SLAC-PUB-\SLACPubNumber \\
March 2003 \\
\end{flushright}

\par\vskip 5cm  

\begin{center}
\Large \bf\boldmath 
Observation of $B$ Meson Decays to $\eta\pi$ and $\eta K$ \\
\end{center}
\bigskip


\begin{center}
\large \bf Abstract
\end{center}
We present preliminary measurements of the $B$ meson decays 
\etapp, \etakp, and \etakz.
The data were recorded with the \babar\ detector 
at PEP II and correspond to 
$88.9\times 10^6$ \BB\ pairs produced in \epem\ annihilation through the
\UfourS\ resonance.  We find the
branching fractions $\Betapip=\Retapip$ and $\Betakp=\Retakp$.
We set a $90\%$ CL upper limit of $\Betakz < \Uetakz$.
The time-integrated charge asymmetries are
$\acp(\etapip)=\Aetapip$ and $\acp(\etakp)=\Aetakp$.

\vfill
\begin{center}
Presented at the XXXVIII$^{th}$ Rencontres de Moriond on\\
QCD and High Energy Hadronic Interactions, \\
3/22---3/29/2003, Les Arcs, Savoie, France
\end{center}

\vspace{1.0cm}
\begin{center}
{\em Stanford Linear Accelerator Center, Stanford University, 
Stanford, CA 94309} \\ \vspace{0.1cm}\hrule\vspace{0.1cm}
Work supported in part by Department of Energy contract DE-AC03-76SF00515.
\end{center}

\newpage
} 

\begin{center}
\small

The \babar\ Collaboration,
\bigskip

%
B.~Aubert,
R.~Barate,
D.~Boutigny,
J.-M.~Gaillard,
A.~Hicheur,
Y.~Karyotakis,
J.~P.~Lees,
P.~Robbe,
V.~Tisserand,
A.~Zghiche
\inst{Laboratoire de Physique des Particules, F-74941 Annecy-le-Vieux, France }
% ...

\end{center}\newpage

\setcounter{footnote}{0}

%
%
%
\section{Introduction}\label{sec:intro}

We report the results of searches for $B$ decays to the charmless
final states\footnote{Except as noted explicitly, we use a particle name
to denote either member of a charge conjugate pair.}
$\eta\pi^+$, $\eta\Kp$, and $\eta\Kz$.  We reconstruct the
$\eta$ mesons in both of the dominant final states $\eta\ra\gamma\gamma\
(\eta_{\gamma\gamma})$ and $\eta\ra\pi^+\pi^-\pi^0\ (\etathrpi)$.
The \Kz\ is reconstructed as $K^0_S\ra\pi^+\pi^-$.
For the charged modes we also measure the direct \CP-violating
time-integrated charge asymmetry, 
$\acp =(\Gamma^--\Gamma^+)/(\Gamma^-+\Gamma^+)$,
where $\Gamma^\pm\equiv\Gamma(B^\pm\ra \eta h^\pm)$.

The interest in these decays was sparked by the first reports of the 
observation of the decay \etapK\ \cite{CLEOetapobs} in 1997.  It had
been pointed out by Lipkin six years earlier \cite{Lipkin} that
interference between two penguin diagrams and the known $\eta/\etapr$
mixing angle conspire to greatly enhance \etapK\ and
suppress \etaK.  Due to a parity flip for the vector \Kstar,
the situation is reversed for the \etapKst\ and \etaKst\ decays.  Though
the general features of this picture have already been borne out by
previous measurements and limits, the details and  possible
contributions of singlet diagrams will only be tested with the measurement
of the branching fraction of all four $(\eta,\etapr)(K,\Kstar)$ decays.

It was pointed out more than 20 years ago (before the discovery of the
$B$ meson) by Bander, Soni and 
Silverman \cite{BSS} that penguin loop diagrams allow substantial
($\gsim20\%$) charge asymmetries in some $B$ decays, and an example
they gave was \etaK.
The necessary ingredients are to have two interfering 
diagrams with different weak and strong phases.  More recently, it was
pointed out that such charge asymmetries can be enhanced in
\etappi\ and \etap\ where the decay rate is small but penguin-tree or
penguin-penguin interference is possible \cite{barshay,dighe}.
A series of quantitative predictions have been made in the past decade with
various factorization approaches \cite{kramer,AKL,yang,beneke}.  There
is general agreement that modes such as \etaK, \etap, and \etappi\ 
are expected to have charge asymmetries of 20\% or larger.  
Most, but not all, of the quantitative calculations predict
negative values for all three decays.
A recent paper \cite{chiang} shows that branching fraction and charge
asymmetry measurements for \etappi\ and \etap\ allow for the determination of 
the strong phase difference between tree and penguin amplitudes and the CKM
angle alpha.

The current knowledge of these decays comes from published measurements from
CLEO~\cite{CLEOetapr} and conference results from \babar\ \cite{BABAReta}
and Belle \cite{Belleeta}.
Table~\ref{tab:OldResults} summarizes these previous results.
We present here analyses incorporating new data.  


\begin{table}[htb]
\caption{Summary of branching fraction results for $B$ decays to $\eta$
    mesons from CLEO~\cite{CLEOetapr}, previous \babar ~\cite{BABAReta} 
    measurements, Belle~\cite{Belleeta}, and the
    present analysis.  The results for all fits are given as well as a 90\%
    CL upper limit if the measured yield is not judged to be significant.
    The overall yields and efficiencies ($\epsilon$) are given as the sum
    of yields and efficiencies from the two $\eta$ decay channels.
}
\label{tab:OldResults}
\begin{center}
\vspace*{0.5cm}
\hspace*{-0.5cm}
\begin{tabular}{l|ccccccc}
    \dbline
Expt.    &\# \BB\ ($\times10^{6}$)&Fit \calB$(\times10^{-6})$&UL \calB$(\times10^{-6})$&Signif. ($\sigma$)&Signal yield&$\epsilon$ (\%) \\
    \sgline
$\etapip$ & & & & & & \\
~~CLEO     &10	& $1.2^{+2.8}_{-1.2}$		& 5.7	&0.6	& $5.7$	& 25.0 \\
~~\babar   &23	& $2.2^{+1.8}_{-1.6}\pm0.1$	& 5.2	&1.5	& $8.0$	& 15.8 \\
~~Belle    &32	& $5.4^{+2.0}_{-1.7}\pm0.6$	& ---  	&4.3	& $15.4$ & 9.5 \\
This result&89&       \retapip          	& ---  	&7.0	& $67.6$ & 16.5 \\
    \sgline
$\etaKp$ & & & & & & \\
~~CLEO     &10	& $2.2^{+2.8}_{-2.2}$		& 6.9 	& 0.8	& $5.9$	 & 24.1 \\
~~\babar   &23	& $3.8^{+1.8}_{-1.5}\pm0.2$	& 6.4 	& 3.7	& $12.9$ & 15.6 \\
~~Belle    &32	& $5.3^{+1.8}_{-1.5}\pm0.6$	& ---	& 4.9	& $16.9$ & 10.6\\
This result&89&      \retakp 			& --- 	& 6.2	& $48.7$ & 17.2 \\
    \sgline
$\etaKz$ & & & & & & \\
~~CLEO     &10	& $0.0^{+3.2}_{-0.0}$		& 9.3 	& 0.0	& $0.0$ & 7.0 \\
~~\babar   &23	& $6.0^{+3.8}_{-2.9}\pm0.4$	& 12.2 	& 3.2	& $5.7$	& 4.2 \\
This result&89&      \retakz 			& 4.6 	& 3.3	& $11.2$ & 5.1 \\
    \sgline
\end{tabular}
\end{center}
\end{table}

%
%
\section{Detector and Data} \label{sec:detector}

The results presented in this paper are based on data collected
in 1999--2002 with the \babar\ detector~\cite{BABARNIM}
at the PEP-II asymmetric $e^+e^-$ collider~\cite{pep}
located at the Stanford Linear Accelerator Center.  An integrated
luminosity of 81.9~fb$^{-1}$, corresponding to 
88.9 million \BB\ pairs, was recorded at the $\Upsilon (4S)$
resonance
(``on-resonance'', center-of-mass energy $\sqrt{s}=10.58\ \gev$).
An additional 9.6~fb$^{-1}$ were taken about 40~MeV below
this energy (``off-resonance'') for the study of continuum backgrounds in
which a light or charm quark pair is produced instead of an \UfourS.

The asymmetric beam configuration in the laboratory frame
provides a boost of $\beta\gamma = 0.56$ to the $\Upsilon(4S)$.
Charged particles are detected and their momenta measured by the
combination of a silicon vertex tracker (SVT), consisting of five layers
of double-sided detectors, and a 40-layer central drift chamber,
both operating in the 1.5-T magnetic field of a solenoid. Photons and
electrons are detected by a CsI(Tl) electromagnetic calorimeter (EMC).

Charged-particle identification (PID) is provided by the average 
energy loss ($dE/dx$) in the tracking devices  and
by an internally reflecting ring-imaging 
Cherenkov detector (DIRC) covering the central region.

%
%
\section{Event Selection} 
\label{sec:presel}

Monte Carlo (MC) simulations \cite{geant}\ of the signal decay modes and
of continuum and \BB\ backgrounds are used to establish the event selection
criteria.  The selection is designed to achieve high efficiency and
retain sidebands sufficient to characterize the background for
subsequent fitting.  Photons must have energy exceeding a threshold
dependent on the combinatoric background of the specific mode:
$E_\gamma>30$ MeV for the two photons used to reconstruct the \piz\
in $\etatothrpi$ candidates, and $E_\gamma>$ 100 MeV for \etatogg .
Additionally, we require that the cosine of the center of mass decay angle for \etagg\ daughters, relative to the flight direction of the $\eta$, have an absolute value of less than 0.95 (0.97) for neutral (charged) decays involving the \etagg .

We select $\etagg$, $\etathrpi$, and $\piz$ candidates with the
following requirements on the invariant mass in \mevcc\ of their final
states: $490< m_{\eta}<600$ for \etagg, $520< m_{\eta} < 570$ for \etathrpi, 
and $120 < m_{\pi^0} < 150$.  For
$\kzs\ra\pip\pim$ candidates we require $488 < m_{K_S} < 508$.  
Typical resolutions are about 16\mevcc\ for \etagg, 4.5\mevcc\ for
\etathrpi, 2.8\mevcc\ for \kzs, and 7\mevcc\ for \piz.
For \kzs\ candidates we require that the three-dimensional flight
distance from the event primary vertex be $>2$ mm, and the
two-dimensional angle between flight and momentum vectors be $<40$ mrad.

We make several particle identification (PID) requirements to ensure the
identity of the signal pions and kaons.
Tracks in \etathrpi\ candidates must have DIRC, $dE/dx$, and EMC
responses consistent with pions.  For the charged \etaKp\ decay, the
prompt charged track must have an associated DIRC Cherenkov angle between
$-5\,\sigma$ and $+2\,\sigma$ from the expected value for a kaon.  
For \etaPip , the DIRC Cherenkov angle must be between
$-2\,\sigma$ and $+5\,\sigma$ from the expected value for a pion.

A $B$ meson candidate is characterized kinematically by the energy-substituted mass
$\mes = \sqrt{(\half s + \pvec_0\cdot \pvec_B)^2/E_0^2 - \pvec_B^2}$ and
energy difference $\DE = E_B^*-\half\sqrt{s}$, where the subscripts $0$ and
$B$ refer to the initial \UfourS\ and to the $B$ candidate, respectively,
and the asterisk denotes the \UfourS\ frame. 
The resolutions on these quantities measured for signal events are 30
MeV and $3.0\ \mevcc$, respectively. 
We require $|\DE|\le0.2$ GeV and $5.2\le\mes\le5.29\ \gevcc$
(the lower limit is $5.22\ \gevcc$ for \fetaggpi).  

\subsection{Tau, QED, and continuum background}

To discriminate against tau-pair and two-photon background, we require
in \etathrpi\ channels that the event contain at least five (four) 
charged tracks for neutral (charged) $B$ pairs.  
In \etagg\ analyses, we require three (two) tracks for neutral (charged)
$B$ pairs. 

To reject continuum background, we make use
of the angle $\theta_T$ between the thrust axis of the $B$ candidate and
that of
the rest of the tracks and neutral clusters in the event, calculated in
the center-of-mass frame.  The distribution of $\cos{\theta_T}$ is
sharply peaked near $\pm1$ for combinations drawn from jet-like $q\bar q$
pairs and is nearly uniform for the isotropic $B$ meson decays; we require
$|\cos{\theta_T}|<0.9$.  A second $B$ candidate satisfying the selection criteria
is found in about 10--20\% of the events.  In this case the ``best''
combination is chosen as the one closest to the nominal $\eta$ mass.

The remaining continuum background dominates the samples and is modeled
from sideband data for the maximum likelihood fits described in
Section~\ref{sec:mlfit}. 

\subsection{\boldmath \BB\ background}

We use Monte Carlo simulations of \BzBzb\ and \BpBm\ pair production and decay 
to look for possible \BB\ backgrounds.  
Most \BB\ backgrounds in these analyses come from other charmless decays.  
From these studies we find no evidence for
significant \BB\ background in the \etatothrpi\ decay chains. 

For the \etatogg\ modes we find potential \BB\ backgrounds from several
charmless final states, which we treat with additional event selection
criteria.  To reduce background from
$\piz\pip$, $\piz \Kp$, and $\piz \Kz$, we eliminate \etagg\ candidates
that share a photon with any \piz\ candidate having
momentum between 1.9 and 3.1 GeV/c in the \UfourS\ frame.  Additionally, we
remove high-energy photons to suppress background from $\Kstar\gamma$,
by requiring $E_{\gamma} < 2.4$ GeV.  We find a small 
remaining \BB\ background in $\etagg \pi$ ($\etagg K$) from
$\eta\rho$ ($\eta\Kstar$).  To discriminate between these and the signal
we include a \BB\ component in the likelihood fits for modes with
\etatogg, as described in Section~\ref{sec:like}.

%
%
\section{Maximum Likelihood Fit}\label{sec:mlfit}

We use an unbinned, multivariate maximum-likelihood fit to extract
signal yields for our modes.  A sample of events to fit is selected
as described in Section~\ref{sec:presel}.

\subsection{Likelihood Function} \label{sec:like}

The likelihood function incorporates four uncorrelated variables.  
We describe the $B$ decay kinematics with two variables: \DE\ and \mes .
We also include $m_{\eta}$ and a 
Fisher discriminant \xf\ which describes energy flow in the event.
The Fisher discriminant combines four
variables: the angles with respect to the beam axis, in the \UfourS\
frame, of the $B$ momentum and $B$ thrust axis, and
the zeroth and second angular moments $L_{0,2}$ of the energy flow
about the $B$ thrust axis.  The moments are defined by
\begin{equation}
  L_j = \sum_i p_i\times\left|\cos\theta_i\right|^j,
\end{equation}
where $\theta_i$ is the angle with respect to the $B$ thrust axis of
track or neutral cluster $i$, $p_i$ is its momentum, and the sum
excludes the $B$ candidate.

As measured  correlations among the observables in the selected data
are small, we take the
probability distribution function (PDF) for each event to be a product 
of the PDFs for the separate observables.  
We define hypotheses $j$, where $j$ 
can be signal, continuum background, or (for modes with \etagg) \BB\
background.
The product PDF (to be evaluated with the
observable set for event $i$) is then given by

\begin{equation}
{\cal P}^i_{j} =  {\cal P}_j (\mes) \cdot {\cal  P}_j (\DE) \cdot
 { \cal P}_j(\xf) \cdot {\cal P}_j (m_{\eta}).
\end{equation}

\noindent The likelihood function for each decay mode is

\begin{equation}
{\cal L} = \frac{\exp{(-\sum_j Y_{j})}}{N!}\prod_i^{N}\sum_j Y_{j} {\cal P}^i_{j}\,,
\end{equation}

\noindent where $Y_{j}$ is the yield of events of hypothesis $j$ found by the
fitter, and $N$ is the number of events in the sample.  The first factor takes 
into account the Poisson fluctuations in the total number of events. 

\subsection{Signal and Background Parameterization}

We determine the PDFs for signal and \BB\ background from
MC distributions in each observable.  For the continuum background we establish
the functional forms and initial parameter values of the PDFs with data
from sidebands in \mes\ or \DE.  We allow several background parameters
to float in the final fit.

The distributions in $m_{\eta}$, and in \mes\ and \DE\ for signal,
are parameterized as
Gaussian functions, with a second or third Gaussian  as
required for good fits to these samples.  Slowly varying distributions
(combinatoric background under the $\eta$ mass and \DE\ peaks) are 
parameterized by linear functions.
The combinatoric background in \mes\ is described by a phase-space-motivated 
empirical function \cite{argus}.  We model the
\xf\ distribution using a Gaussian function with different widths above 
and below the mean, and include a
linear contribution of 1--3\%\ in area to account for outlying events.
The linear term ensures that the significance of 
the signal is not overestimated relative to background.  
Because of the rarity of outlying events this component is not
particularly well determined in some data samples, but we have checked
that the yield and its significance are
insensitive to choices of a linear component over the conservative range
1--6\%. 

We check the simulation on which we rely for signal PDFs by comparing
with large data control samples.  For \mes\ and \DE\ we use
the decays \Dcontrol, which have similar topology to the modes under study.
For $m_{\eta}$ we use inclusive resonance production.

%
%
\section{Fit Results}

By generating (from PDF shapes) and fitting simulated samples of signal
and background, we verify that our fitting procedure is functioning
properly.  We find that the minimum $\ln{\cal L}$ value in the
on-resonance sample lies well within the $\ln{\cal L}$ distribution
from these simulated samples.

The efficiency is obtained from the fraction of signal MC events passing
the selection, adjusted for any bias in the likelihood fit.  This bias is
determined from fits to simulated samples, each equal in size to the
data and containing a known number of signal MC events combined with events
generated from the background PDFs.  We find biases ranging from 1\% to
4\%, depending on the mode.

\providecommand{\corrEffB}{Corr. $\epsilon$$\times$$\prod\calB_i$ (\%)}
\providecommand{\signifSyst}{Signif. w syst. ($\sigma$)}
\providecommand{\bfemsix}{${\cal B}(\times10^{-6}$)}
\begin{table}[htbp]
\caption{
Final fit results for \etahp\ and \etaKz\ , where \etatothrpi\ and \etatogg .
We report branching fractions for the two $\eta$ decay channels separately 
($\calB$) and after combining the results of the two channels 
(Combined $\calB$).  Systematic contributions are included in
the signficance values. The Corrected $\calB$ for the charged modes is the 
branching fraction after correcting for crossfeed from one charged mode into 
the other.
}
\label{tab:etahksresults}
\begin{center}
\vspace*{-0.2cm}
\hspace*{-1.0cm}
\begin{tabular}{lcccccc}
\dbline
Fit quantity 	&\fetathrpipi\  &\fetaggpi      &\fetathrpik\		&\fetaggk	&\fetathrpikz	&\fetaggkz  \\
\sgline			                                 					
Fit sample size     	&               &               & 			&		& 		&		\\
~~On-resonance    	& 9477          & 6933          & 5383			& 5884		& 1270		& 1435		\\
~~Off-resonance   	& 1104          & 1168          & 630			& 959		& 158		& 183		\\
Signal yield		&               &               &			&		&		& 		\\
~~On-res data		&$28.0^{+10.0}_{-8.8}$  &$39.6^{+11.3}_{-10.1}$ &$14.4^{+8.2}_{-7.0}$	&$34.3^{+9.8}_{-8.8}$	&$2.6^{+4.1}_{-3.1}$	&$8.6^{+4.8}_{-3.8}$ \\
~~Off-res data		&$1.1^{+2.1}_{-1.2}$    &$0.0^{+2.3}_{-0.0}$    &$0.6^{+3.9}_{-2.9}$	&$0.0^{+0.7}_{-0.0}$	&$0.0^{+0.7}_{-0.0}$	&$0.0^{+0.8}_{-0.0}$ \\
Selection $\epsilon$ (\%)& 23.3         & 28.7		& 22.6			& 30.6		& 22.6		& 24.8		\\
$\prod\calB_i$ (\%)	 & 22.6         & 39.4		& 22.6			& 39.4		& 7.8		& 13.5		\\
$\epsilon\times\prod\calB_i$ (\%) & 5.2	& 11.3		& 5.1			& 12.1		& 1.76 		& 3.34		\\ 
Stat. sign. ($\sigma$)   & 4.3           & 5.7		& 2.4			& 5.7		& 0.8  		& 3.2		\\ 
\sgline			 
${\cal B}(\times 10^{-6})$	&$6.0^{+2.1}_{-1.9}$  &$3.9^{+1.1}_{-1.0}$  &$3.2^{+1.8}_{-1.5}$  &$3.2^{+0.9}_{-0.8}$  &$1.7^{+2.6}_{-2.0}$ &$2.9^{+1.6}_{-1.3}$\\
~~Combined ${\cal B}$	& \multicolumn{2}{c}{$4.5^{+1.0}_{-0.9}\pm 0.3$}
			& \multicolumn{2}{c}{$3.2^{+0.8}_{-0.7}\pm 0.2$}
			& \multicolumn{2}{c}{\retakz } \\
~~Stat. sign. ($\sigma$)   & \multicolumn{2}{c}{7.0}        &  \multicolumn{2}{c}{6.2} 		 		& \multicolumn{2}{c}{3.3}	\\ 
~~Corrected ${\cal B}$	& \multicolumn{2}{c}{\retapip}
			& \multicolumn{2}{c}{\retakp}
			& \multicolumn{2}{c}{\retakz} \\
90\% C.L. UL(incl. syst.) & \multicolumn{2}{c}{---}
			& \multicolumn{2}{c}{---}
			& \multicolumn{2}{c}{$<\uetakz$} \\
\sgline
Bkg \acp 	& $-0.00\pm0.01$ & $-0.02\pm0.01$ & $-0.02\pm0.01$ & $-0.00\pm0.01$ & --		& --		\\
Signal \acp	& $-0.50\pm0.31$ & $-0.51\pm0.24$ & $-0.56\pm0.55$ & $-0.25\pm0.26$ & --		& --		\\
~~Combined \acp & \multicolumn{2}{c}{\Aetapip}		& \multicolumn{2}{c}{\Aetakp}	& \multicolumn{2}{c}{--}	\\
~~Stat. sign. ($\sigma$)   & \multicolumn{2}{c}{2.5}        &  \multicolumn{2}{c}{1.4} 		 		&   		& 		\\ 
\dbline
\end{tabular}
\end{center}
\end{table}

In Table~\ref{tab:etahksresults} we show the results of the fits for
off- and on-resonance data.  
Shown for each decay mode are the number of events that were fit, the
signal yield, the efficiency ($\epsilon$) and daughter branching
fraction product ($\prod\calB_i$), and the central 
value of the branching fraction. 
We also show the branching fraction results after combining the two 
$\eta$ decay channels, before and after a correction for crossfeed
between the two charged channels (see Section~\ref{sec:combine}), and the 
statistical significance of this combined result.
For \fetakz\ we quote a 90\% CL upper limit.
The statistical error on the number of events
is taken as the change in the central value when the quantity
$-2\ln{\cal L}$ changes by one unit. The statistical significance is
taken as the square root of the difference between the value of
$-2\ln{\cal L}$ for zero signal and the value at its minimum.  
For the charged modes we also give the charge asymmetry \acp .

\begin{figure}[htbp]
\vspace{0.5cm}
 \includegraphics[angle=0,width=\linewidth]{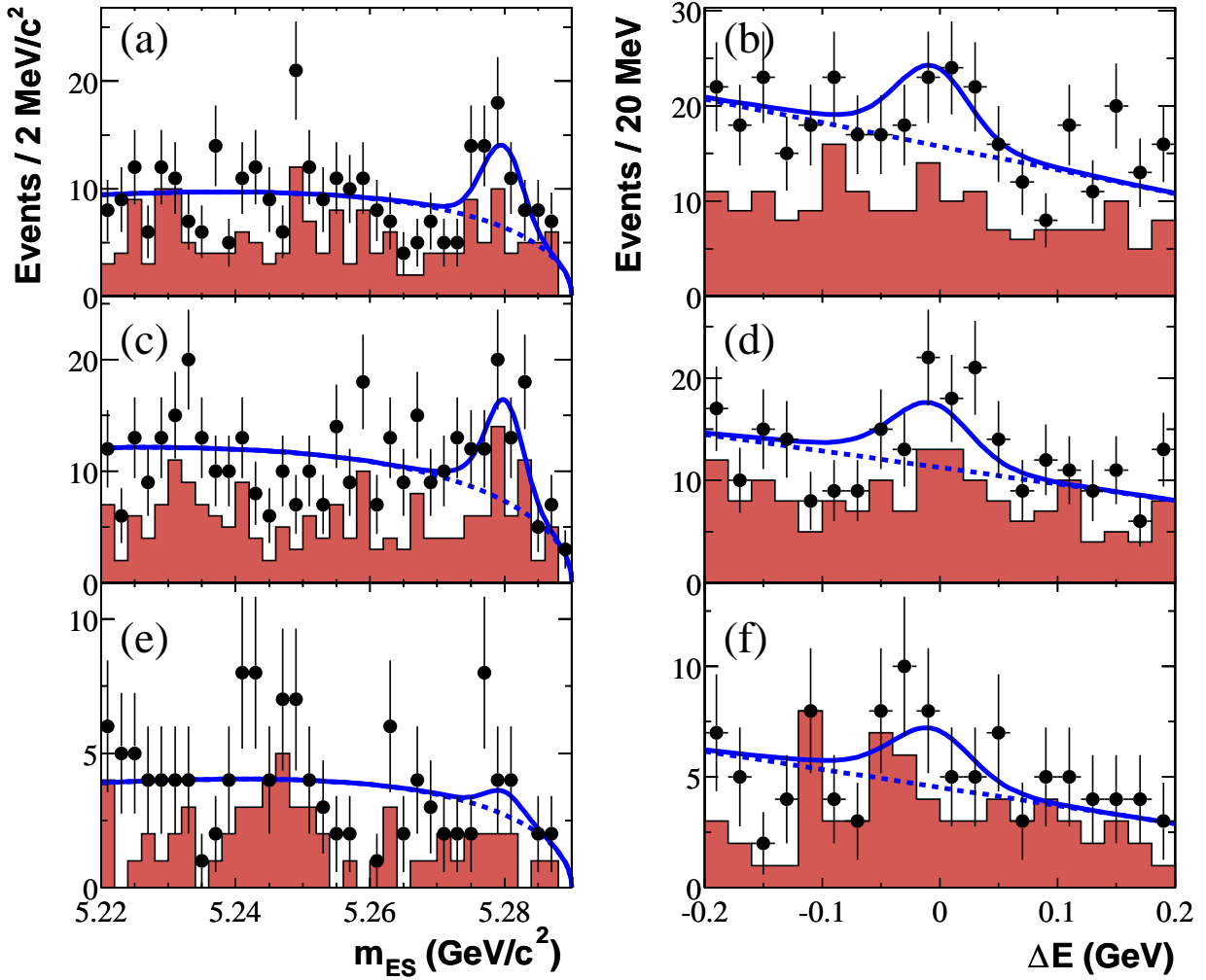}
 \caption{\label{fig:projMbDE}
Projections of the $B$ candidate \mes\ and \DE\ for \etapp\ (a, b), \etakp\ (c, d), and
\etakz\ (e, f). Points with errors represent data, shaded histograms the
 \etagg\ subset, solid curves the full fit functions,
 and dashed curves the background functions.
These plots are made with a cut on the signal likelihood and thus do not 
show all events in the data samples.
  }
\end{figure}

In Fig.\ \ref{fig:projMbDE}\ we show projections of \mes\ and \DE\ made by
selecting events with signal likelihood (computed without the variable
shown in the figure) exceeding a mode-dependent threshold that optimizes the
expected sensitivity.

%
%
\section{Systematic Uncertainties}
\label{sec:syst}

Most of the systematic errors on yields that arise from uncertainties in the 
values of the PDF parameters have already been incorporated into the overall
statistical error, because their background parameters are free in the
fit.  We determine the sensitivity to parameters of the signal PDF
components by varying these within their uncertainties.  The results are
shown in the first row of Table \ref{tab:systtab}.  This is the only
systematic error on the fit yield; the other systematics apply to either
the efficiency or the number of \BB 's.

The uncertainty in our knowledge of the efficiency is found 
to be 0.8$N_t$\%, 2.5$N_\gamma$\%, and 3\%\ for a
\KS\ decay, where $N_t$ and $N_\gamma$ are the number of signal tracks
and photons, respectively.  We estimate the uncertainty in the number of 
produced \BB\ pairs to be 1.1\%.  The estimate
of systematic bias from the fitter itself (1--2\%) comes from fits of
simulated samples with varying background populations.  Published world
averages \cite{PDG2002}\ provide the $B$ daughter branching fraction
uncertainties. 
We account for systematic effects in \costhr\ (1\%) and in the PID
requirement (0.5\%) on the prompt charged track.  Values for each of
these contributions are given in Table \ref{tab:systtab}. 

A study of the charge asymmetry as a function of momentum for all tracks in 
hadronic events bounds the tracking efficiency component of charge-asymmetry 
bias to be less than 1\%.  Samples of $B$ and $D^*$-tagged $D\ra K\pi$
decays provide additional crosschecks that the bias is small.
We assign a systematic uncertainty for \acp\ of 1.1\% based on the tracking 
study and a small PID contribution determined from the $D^*$ studies.

\begin{table}[htbp]
\caption{Estimates of systematic errors (in percent) for the \etahp\ and \etaKz\  modes.  We specify which systematics are uncorrelated (U) or correlated (C) 
between $\eta$ decay channels.}
\label{tab:systtab}
\begin{center}
\begin{tabular}{l|cc|cc|cc}
\dbline
Quantity & \fetathrpipi\ & \fetaggpi	& \fetathrpik 	& \fetaggk  &\fetathrpikz & \fetaggkz \\
\sgline
Fit yield (U)             & 3.9 & 3.7 & 8.4 & 4.5 & 20.7 & 2.3  \\
Fit efficiency/bias (U)   & 1.9 & 1.3 & 1.3 & 0.8 & 1.0  & 1.7  \\
Track multiplicity (C)    & 1.0 & 1.0 & 1.0 & 1.0 & 1.0  & 1.0  \\
Tracking eff/qual (C)     & 2.4 & 0.8 & 2.4 & 0.8 & 3.7  & 2.1  \\
\piz/$\eta/\gamma$ eff (C)& 5.0 & 5.0 & 5.0 & 5.0 & 5.0  & 5.0  \\
\KS\ efficiency (C)       & --- & --- & --- & --- & 2.9  & 2.9  \\
Number \BB\ (C)           & 1.1 & 1.1 & 1.1 & 1.1 & 1.1  & 1.1  \\
Branching fractions (U)   & 1.0 & 1.0 & 1.0 & 1.0 & 1.0  & 1.0  \\
MC statistics (U)         & 1.1 & 1.1 & 1.0 & 1.1 & 1.1  & 1.0  \\
\costhr (C)               & 1.0 & 1.0 & 1.0 & 1.0 & 1.0  & 1.0  \\
PID (C)                   & 1.4 & 1.0 & 1.4 & 1.0 & 1.0  & ---  \\
\sgline
Total                     & 7.5 & 6.9 & 10.5 & 7.3 & 22.0 & 7.2  \\
Uncorrelated              & 4.6 & 4.2 &  8.6 & 4.8 & 20.8 & 3.2  \\
Correlated                & 6.0 & 5.5 &  6.0 & 5.5 &  7.2 & 6.4  \\
\dbline
\end{tabular}
\end{center}
\end{table}

\section{\boldmath Combined Results}
\label{sec:combine}

\begin{figure}[htbp]
 \psfiletwoBB{54 145 529 610}{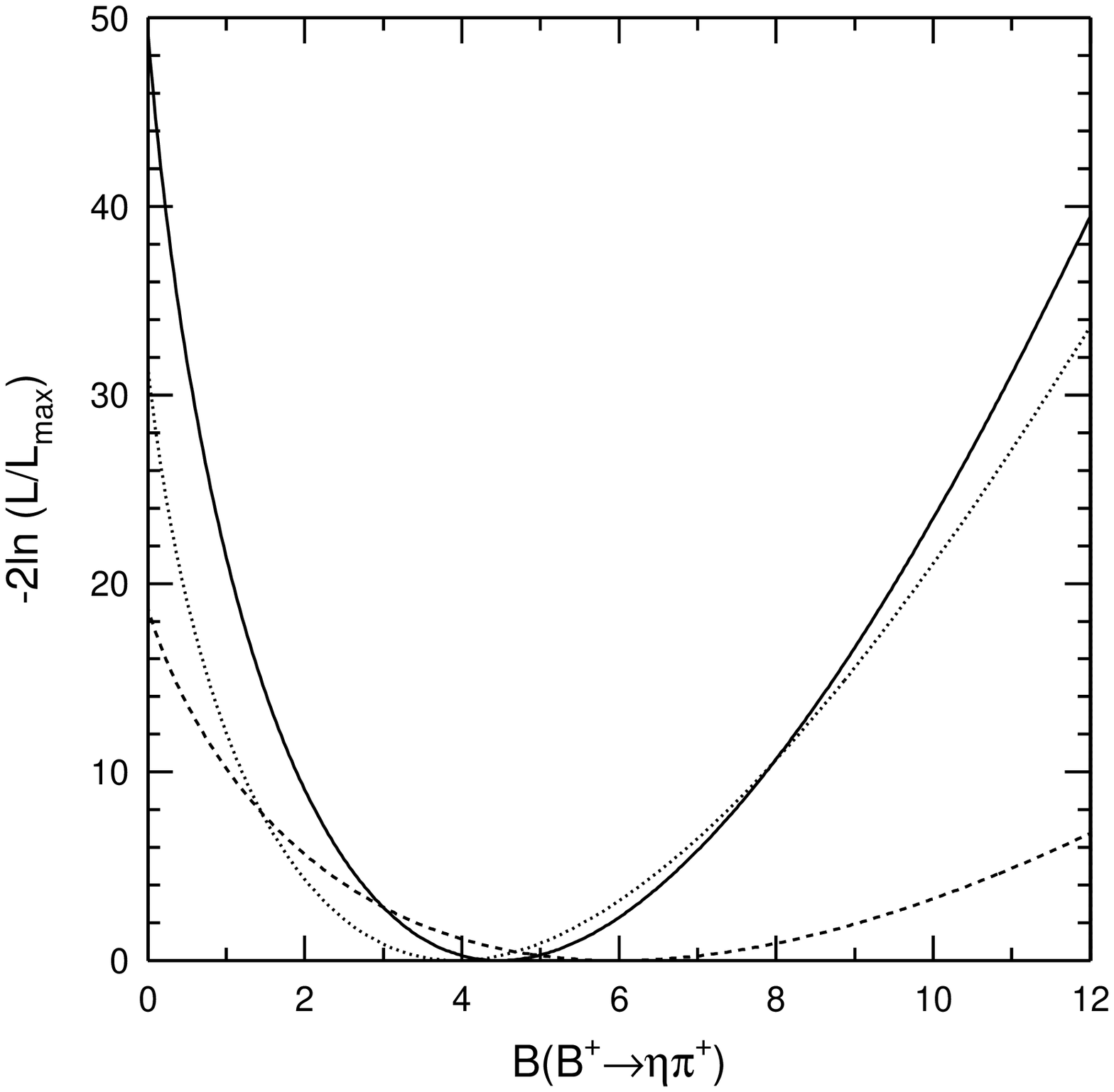}%
	     {54 145 529 610}{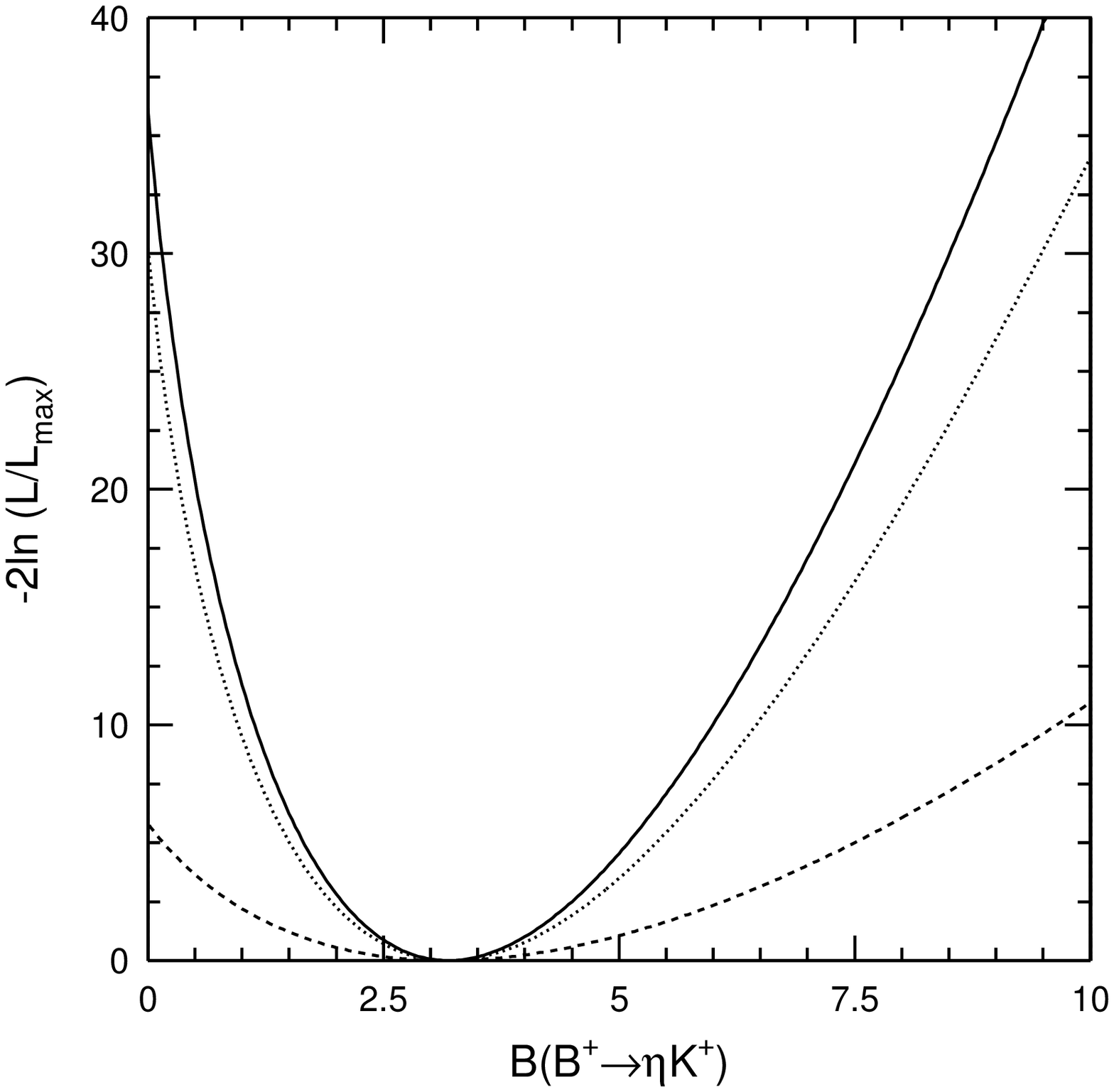}{1.0}
 \psfile[54 145 529 610]{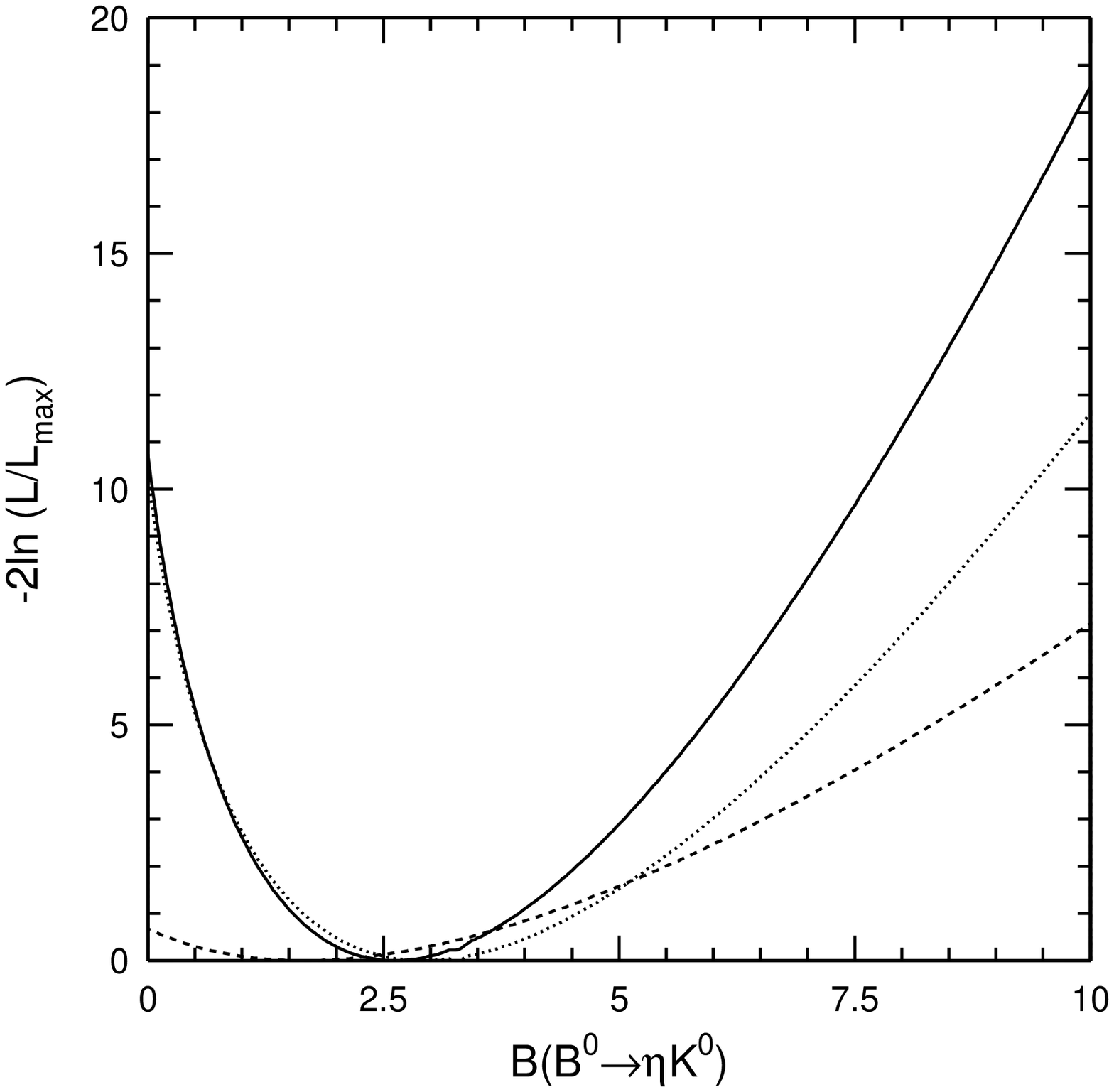}{.5}
 \vspace{-0.7cm}
 \caption{\label{fig:etahchi}
Distributions of $-2\ln{\cal L}$ vs branching fraction for \fetapip ,
\fetakp\ and \fetakz\ 
decays.  Two secondary channels (dashed and dotted lines) are combined
to produce a final result (solid line).  The dashed line corresponds to
\etatothrpi\ decays, while the dotted line corresponds to \etatogg .
}
\end{figure}

\begin{figure}[htbp]
 \psfiletwoBB{54 145 529 610}{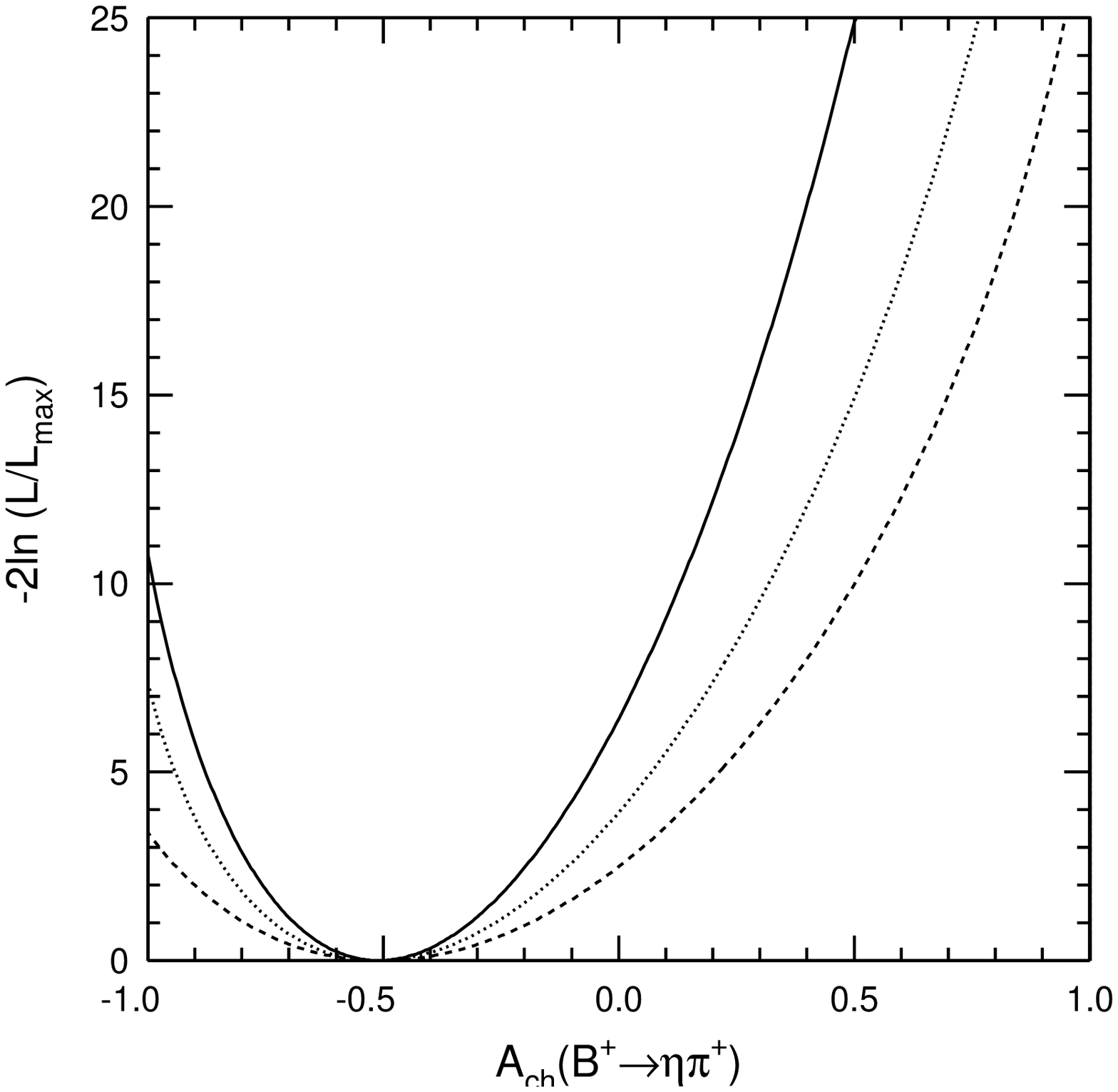}%
	     {54 145 529 610}{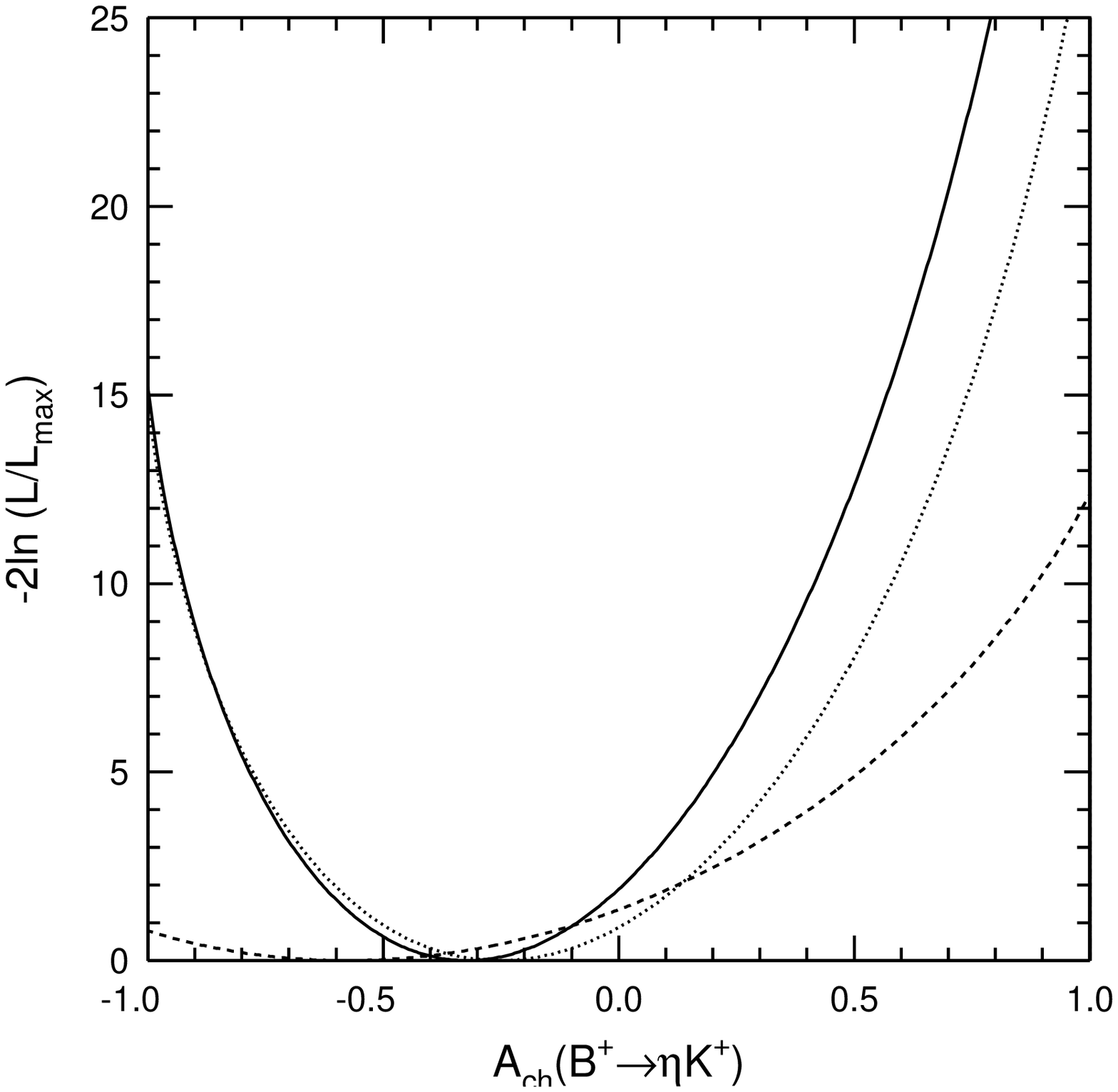}{1.0}
 \vspace{-0.7cm}
 \caption{\label{fig:etahchiAch}
Distributions of $-2\ln{\cal L}$ vs \acp\ for \fetapip and \fetakp\ 
decays.  Two secondary channels (dashed and dotted lines) are combined
to produce a final result (solid line).  Dashed line corresponds to
\etatothrpi\ decays, while the dotted line corresponds to \etatogg .
}
\end{figure}

We next combine the $\eta\ra3\pi$ and $\eta\ra\gamma\gamma$ branching fraction
measurements.  We do this
by first forming for each $\eta$ decay mode the convolution of \calL\
from the fit with a Gaussian function representing the uncorrelated 
systematic error.  The curves $-2\ln{\calL}$
are shown in Fig. \ref{fig:etahchi}, for each $\eta$ mode and for their sum.
For the time integrated charge asymmetries the corresponding $-2\ln{\calL}$
plots are given in Fig.\ \ref{fig:etahchiAch}.

The results at this stage are given in the row labeled ``Combined ${\cal
B}$'' in Table \ref{tab:etahksresults}.
For the charged modes we must apply a correction for kaon--pion 
crossfeed arising from imperfect PID.  
In studies with kaon and pion samples tagged kinematically from the
decays $D^{*+}\ra\pi^+D^0$, $D^0\ra K^-\pi^+$ we find that $9\pm2\%$ of
pions are accepted by our kaon selection and vice versa.  After
correcting for this and adding the associated systematic uncertainty we
obtain the final measurements summarized in
Section~\ref{sec:conclusion}.

%
%
\section{Conclusion}
\label{sec:conclusion}

We report preliminary measurements of branching fractions and \acp\ for $B$ meson decays to
$\eta$ with a charged kaon or pion, as well as the branching fraction
for \etaKz.  
We find statistically significant signals in the charged $B$ decays. 
The branching fractions are
\begin{eqnarray*}
\Betapp &=& \Retapip\,, \\
\Betakp &=& \Retakp. 
\end{eqnarray*}

\noindent For the neutral $B$ decay, we find $\Betakz=\Retakz$.  Since
the statistical significance of this result is only 3.3$\sigma$, we
determine a 90\% CL upper limit:
\begin{eqnarray*}
\Betakz & < & \Uetakz.
\end{eqnarray*}

\noindent 
These results supersede the previous \babar\ measurements~\cite{BABAReta}.
Our measurements of the \CP-violating charge asymmetries in the charged
modes are
\begin{eqnarray*}
\acp(\etapip) &=& \Aetapip\,, \\
\acp(\etakp) &=& \Aetakp.
\end{eqnarray*}
These charge asymmetry
results are in agreement with the theoretical expectations discussed
in Section \ref{sec:intro}\ and rule out substantial positive asymmetries.

%

\section{Acknowledgments}
\label{sec:Acknowledgments}


We are grateful for the 
extraordinary contributions of our \pep2\ colleagues in
achieving the excellent luminosity and machine conditions
that have made this work possible.
The success of this project also relies critically on the 
expertise and dedication of the computing organizations that 
support \babar.
The collaborating institutions wish to thank 
SLAC for its support and the kind hospitality extended to them. 
This work is supported by the
US Department of Energy
and National Science Foundation, the
Natural Sciences and Engineering Research Council (Canada),
Institute of High Energy Physics (China), the
Commissariat \`a l'Energie Atomique and
Institut National de Physique Nucl\'eaire et de Physique des Particules
(France), the
Bundesministerium f\"ur Bildung und Forschung and
Deutsche Forschungsgemeinschaft
(Germany), the
Istituto Nazionale di Fisica Nucleare (Italy),
the Foundation for Fundamental Research on Matter (The Netherlands),
the Research Council of Norway, the
Ministry of Science and Technology of the Russian Federation, and the
Particle Physics and Astronomy Research Council (United Kingdom). 
Individuals have received support from 
the A. P. Sloan Foundation, 
the Research Corporation,
and the Alexander von Humboldt Foundation.


\begin{thebibliography}{99}

\bibitem{CLEOetapobs}
CLEO Collaboration, B. H. Behrens \etal, \jprl{80}, 3710 (1998).

\bibitem{Lipkin}
H.\ J.\ Lipkin, \plb{254}, 247 (1991).

\bibitem{BSS}
M. Bander, D. Silverman, and A. Soni, \jprl{43}, 242 (1979).

\bibitem{barshay}
S. Barshay, D. Rein, and L.M. Sehgal, \plb{259}, 475 (1991).

\bibitem{dighe}
A.S. Dighe, M. Gronau, and J.L. Rosner, \jprl{79}, 4333 (1997).

\bibitem{kramer}
G. Kramer, W.F. Palmer, and H. Simma, \npb{428}, 77 (1994).

\bibitem{AKL}
A. Ali, G. Kramer, and C.-D. L\"{u}, \jprd{59}, 014005 (1999).  These
authors use the opposite sign convention for \acp\ than the one used in
this paper.

\bibitem{yang}
M.-Z. Yang and Y.-D. Yang, \npb{609}, 469 (2001).

\bibitem{beneke}
M. Beneke and M. Neubert, \npb{651}, 225 (2003).

\bibitem{chiang}
C.-W. Chiang and J. L. Rosner, \jprd{65}, 074035 (2002).

\bibitem{CLEOetapr}
CLEO Collaboration, S.\ J.\ Richichi \etal, \jprl{85}, 520 (2000).

\bibitem{BABAReta}
P. Bloom, Proceedings of the 2002 SLAC Summer Institute, hep-ex/0302030 (2003).

\bibitem{Belleeta}
H.C. Huang (for the Belle Collaboration), hep-ex/0205062, Moriond 2002 
contributed paper (2002).

\bibitem{BABARNIM}
\babar\ Collaboration, B.\ Aubert \etal, \nima{479}, 1 (2002).

\bibitem{pep} 
PEP-II Conceptual Design Report, SLAC-R-418 (1993).

\bibitem{geant}
The \babar\ detector Monte Carlo simulation is based on GEANT:
S. Agostinelli \etal, CERN-IT-20020003, KEK Preprint 2002-85, SLAC-PUB-9350,
    submitted to \nima{}.

\bibitem{argus}
With $x\equiv\mes/E_b$ and $\xi$ a parameter to be fit, $f(x) \propto
x\sqrt{1-x^2}\exp{\left[-\xi(1-x^2)\right]}$.  See ARGUS Collaboration,
H.\ Albrecht \etal, \plb{241}, 278 (1990).

\bibitem{PDG2002}
Particle Data Group, 
K.~Hagiwara \etal, \jprd{66}, 010001 (2002).

\end{thebibliography}
\end{document}